- **Advantages from Quantum Principle**: Utilizes a quantum-entangled photon source, offering innovative microscopy based on quantum non-locality over classical microscopic techniques.
- **Improved Image Contrast and Sensitivity**: Taking the second-order correlation between the quantum entanglement photon pair enhances the contrast of image and sensitivity compared to the classical polarization microscopy.
- **Robust and Alignment Free Imaging**: Based on the quantum interference technique, high resistance against stray light or perturbation from surrounding medium, and unnecessary for beam alignment, adjustment of path difference (delay line) or canceling a phase drift.
- **Low Illumination Power:** Constructs images with extremely low illumination power (on the order of pico-joule), which minimizes sample damage and ensures energy efficiency during image construction.

# Demonstration of Quantum Polarization Microscopy using an Entangled-Photon Source

*Mousume Samad [1,2], Maki Shimizu [1], Yasuto Hijikata [1]*

1. Graduate School of Science and Engineering, Saitama University, 255 Shimo Okubo, Saitama-shi, Saitama 338- 8570, Japan.
2. Department of Information and Communication Engineering, Bangladesh Army University of Engineering and Technology (BAUET), Qadirabad 6431, Bangladesh.

ABSTRACT

With the advancement of non-classical light sources such as single-photon and entangled-photon sources, innovative microscopy based on the quantum principles has been proposed over traditional microscopy. This paper introduces the experimental demonstration of a quantum polarization microscopic technique that incorporates a quantum-entangled photon source. Although the point that employs the variation of polarization angle due to reflection or transmission at the sample is similar to classical polarization microscopy, the method for constructing image contrast is significantly different. Image contrast is constructed by the coincidence count of signal and idler photons. In the case that coincidence count is recorded from both the signal and idler photons, the photon statistics resemble a thermal state, similar to the blackbody radiation, but with a significantly higher peak intensity in the second order autocorrelation function at zero delay that is derived from coincidence count. While, when the coincidence count is taken from either the signal or idler photon only, though the photon state exhibits a thermal state again, the photon statistics become more dispersive and result in a lower peak intensity of the autocorrelation function. These different thermal states can be switched by slightly changing the photon polarization, which is suddenly aroused within narrow range of analyzer angle. This polarization microscopic technique can provide a superior imaging technique compared to the classical method, opening a new frontier for research in material sciences, biology, and other fields requiring high-resolution imaging.

**1. Introduction**

Optical microscopy and spectroscopy play significant roles in modern research across a wide range of disciplines, including fundamental physics, materials science, chemical studies, and biological sciences. In optical microscopy, it is fascinating to observe how historical advancements in light properties have consistently led to the creation of innovative imaging applications. Optical imaging with classical light is limited in terms of signal-to-noise ratio, resolution, contrast, and spectral range [1]. To overcome these limitations quantum properties of light utilizing correlated, entangled, or squeezed photons have been innovated for optical imaging [2-6]. In addition to this, quantum imaging might serve the extremely robust and highly resistive measurements against various kinds of perturbations such as mechanical vibration, fluctuation of surrounding circumstances or light source, and so on. The initial demonstration of entangled-photon sources, known as the bi-photon state, has led to numerous applications in quantum imaging [7-9]. In particular, the essential features of entangled-photon sources are polarization state, momentum, energy, and position correlations of entangled-photon pair [10-11]. The entangled-photon source can enable the imaging technique in spectral ranges where efficient detection is challenging, or even imaging with light that does not directly interact with the sample [12-15]. Moreover, utilizing specific quantum states of light and their photon statistics allows for sensing and imaging beyond the constraints of classical techniques. The quantum optical microscopy with entangled-photon source can construct an image with extremely low light intensity when applying quantum light, enabling new insights into photosensitive biochemical applications.

At present polarization-entangled-photon sources have been a key for these researches due to their simplicities in generation, control, and measurement. The entangled-photon pairs from these sources can be produced through spontaneous parametric down-conversion (SPDC) [16]. In classical imaging methods, the key technology is extracting the phase information by interfering with a reference beam. It is impossible to extract phase information from classical interference measurements when incoherent and unpolarized beams are used. Whereas the coherence of the polarization-entangled state allows for the extraction of images despite dynamic phase disorder and significant classical noise [5]. In optical microscopy, there are some challenges for biological measurement such as the need to increase power to improve resolution while also decreasing power to prevent photo damage to the biological structure. This creates a trade-off between resolution and photo damage. Moreover, the low illumination power leads to the reduction in the resistivity against noise. In contrast, quantum polarization microscopy with an entangled-photon source can be carried out at extremely low light intensity to enable new insights into photosensitive biomedical applications.

An entangled-photon source microscopy technique proposed in Ref. [17] represents the first demonstration of entangled-photon source optical phase measurement that is beyond the standard quantum limit (SQL) for imaging. While improving the signal-to-noise ratio (SNR) of phase measurements beyond the standard quantum limit (SQL) is crucial, entanglement is necessary for achieving this enhancement [18-19]. Additionally, quantum imaging utilizing spatial and polarization entanglement for biological organisms is described in Ref. [20]. This approach, known as imaging by coincidence from entanglement (ICE), provides high SNR, a greater resolvable pixel count, and a full birefringence-quantified quantum imaging technique that produces high-quality images of biological samples. Quantum microscopy at the Heisenberg limit, proposed in Ref. [21] involves coincidence quantum microscopy (QMC) with balanced path lengths, enabling super-resolution imaging with substantially higher speeds and contrast-to-noise ratios (CNRs) compared to existing wide-field quantum imaging methods. Despite these advantages, these techniques are significantly affected by low SPDC efficiency or detector limits. For example, in the case of quantum holography using polarization entanglement [5], it takes the order of a few 10 hours for image acquisition due to the low frame rate of EMCCD cameras.

In this paper, to overcome the issues above, we propose an imaging technique involving measurements of coincidence counts and the autocorrelation function at zero delay, $g^2(0)$, using a quantum polarized type II entangled-photon source. According to our method, the variation of polarization angle due to reflection or transmission at the sample is detected using two distinct thermal states of SPDC photon statistics. Namely, since it switches between the two different states depending on whether either the signal or idler port is in vacuum state or not, by setting the analyzer at the sample side to block the photon, image contrast is constructed due to the transition between the states. It should be also noticed that our method has advantage in which we do not need any precise optical setup such as beam alignment, adjustment of path difference (delay line) or cancelation of a phase drift.

## 2. Methodology of imaging

The entangled-photon pair source used in this experiment exhibits bunching characteristics, a phenomenon associated with the photon statistical properties of light. This behavior, where photons arrive in pairs or "bunches" rather than independently, is typical in SPDC and confirms the photon-pair generation. Actually, the bunching effect is observed in the second-order autocorrelation measurement, $g^2(\tau)$, between signal and idler photons [22]. In general, thermal state exhibits bunching characteristic, as shown by the autocorrelation at zero delay larger than 1 ($g^2(0)$ >1.) However, different from the case of ordinary thermal state like a black-body radiation, a higher bunching peak is seen in the case of SPDC photon statistics because the time interval between photon pairs [Signal and Idler] is longer compared to the Poisson case ($g^2(0)=1$) or super-Poisson ($g^2(0)>1$) [23], leading to the enhancement of photon-bunching. Here, this photon state is called as 'hyper-Poisson state.' Fig. 1(a) shows the schematically illustration of the variations of $g^2(\tau)$ with respect to the time delay ($\tau$) in the cases of hyper-Poisson and super-Poisson states. On the other hand, when either the signal or the idler photon is blocked using an analyzer (or polarizer), a bunching peak with lower intensity is seen as the super-Poisson state. This is due to the time interval tends to cluster together more than in a Poisson distribution [23] and the photon number for each photon pulse becomes more dispersive than in the case of a hyper-Poisson state. This will be experimentally confirmed in the next section. It is noticed that the coincidence window for measurements should be sufficiently smaller than the width of the bunching peak ($\Delta\tau$) to observe the bunching peak correctly. The transition between the hyper-Poisson and super-Poisson state occurs when either the signal or idler photons are blocked/transmitted by the analyzer introduced into the beam path. This transition happens rapidly and smoothly, even with minimal perturbation, resulting in a sharp dip in the $g^2(0)$ curve, as shown in Fig. 1(b).

In this research, optical imaging is constructed through the coincidence count rate or the autocorrelation function at zero delay $g^2(0)$ between signal and idler photon with respect to the polarizer angle. These functions are schematically illustrated in Fig. 1(b). The coincidence count rate mostly follows a sinusoidal curve, as does the transmitted light intensity when the polarizer is rotated. When the coincidence rate approaches zero, which induces the switching from the hyper-Poisson state to the super-Poisson state, the $g^2(0)$ curves indicates a sharp dip as shown in the figure. Therefore, these count rate variations can be used for constructing an image. Conversely, in the polarizer angle range where these count rates are almost constant such as the flat regions in $g^2(0)$ or the coincidence rate at minimum or maximum points, contrast is hardly obtained. Since the sharpness of this dip correlates directly with the image contrast, the autocorrelation imaging may have the capability to provide a higher contrast than that of a coincidence rate.

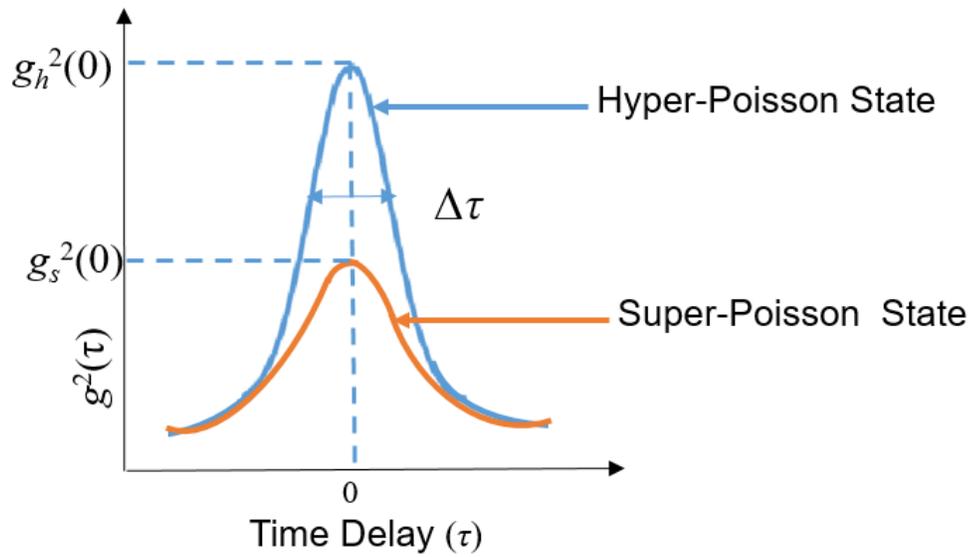

(a)

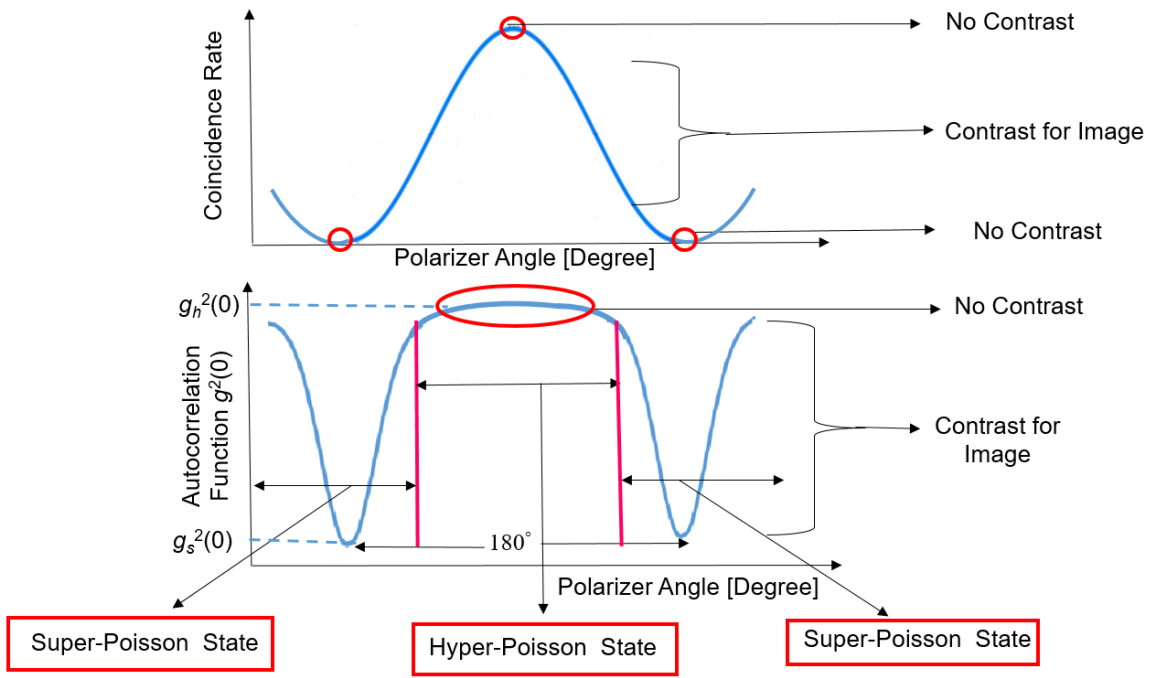

(b)

**Fig.1.** Concept of the imaging methodology: (a) Second-order autocorrelation function $g^2(\tau)$ for hyper-Poisson and super-Poisson states; (b) Coincidence rate and autocorrelation function at zero delay ($g^2(0)$) as a function of polarization angle.

As mentioned above, since the coincidence rate varies almost sinusoidal with respect to the polarization angle, the width of peak (dip) in the curve is π/4 radians, which corresponds to the dynamic range of the measurement. While the width of the autocorrelation function is attributed to fluctuations in the SPDC process. In an entangled-photon source based on SPDC, the time interval of the entangled-photon pair varies slightly due to inherent statistical fluctuations in the emission process. These fluctuations result in a spread in the photon's arrival times at the detectors, corresponding to the width of the dips in $g^2(0)$ plots.

## 3. Experimental set up of the system:

In the present work, the schematic arrangement of quantum polarized microscopy is depicted in Fig. 2. The correlated photon pair source with a collinear type-II spontaneous parametric down-conversion (SPDC), operating wavelength (810 ± 2 nm), and maximum photon pairs per second (>450 kHz) were used as the entangled-photon source. In this system, the signal photons are vertically polarized, while the idler photons are horizontally polarized, and both are transmitted through polarization-maintaining single-mode fibers. Two analyzers are positioned in front of the photon counting detector, one remains fixed, while the other can be rotated, followed by a band-pass filter (810 ± 5 nm) used to minimize stray light. The signal and idler photons were detected using avalanche photodiode (APD) detectors connected via multimode optical fibers. The photon counting of each detector, coincidence counts and delay adjustment between two detectors, were performed by a time correlator.

To demonstrate the proposed imaging method, a first-order (*m*=1) optical spiral retarder sample has been used in this experiment. This sample converts a linearly polarized source into radially or azimuthally polarization. Fig. 3 illustrates the phenomenon where a linearly polarized beam is incident on the first-order optical spiral retarder. When a linear polarization beam is incident (Fig. 3(a)), since the fast axis orientation of the sample is designed to rotate by half angle of azimuth angle, as shown in Fig. 3(b), the output polarization is aligned radially (Fig. 3(c)).

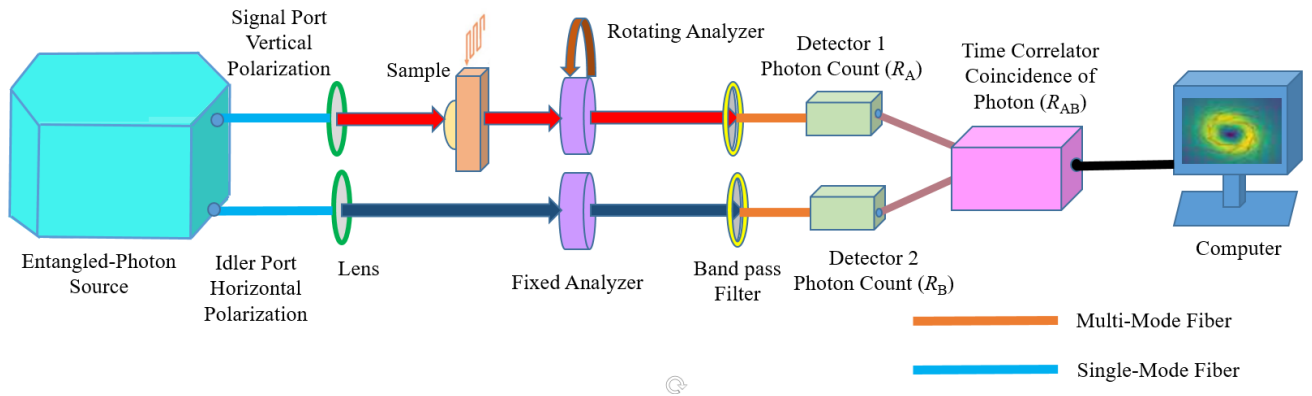

**Fig. 2.** Schematic view of the imaging system.

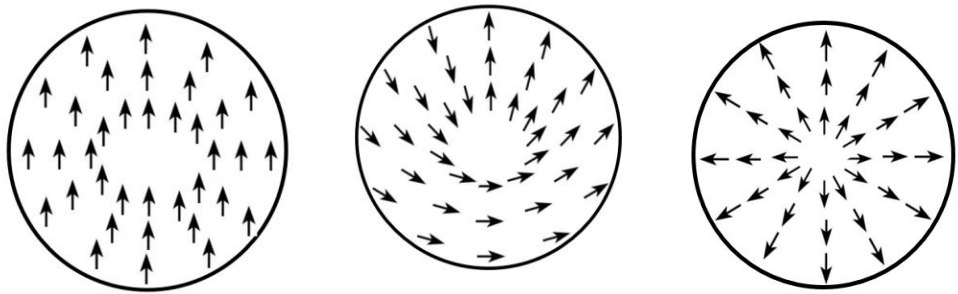

**Fig. 3.** Schematic illustrations for the optical spiral retarder: (a) Linearly polarized input beam incident on optical spiral retarder; (b) Fast axis orientation of optical spiral retarder; (c) Output polarization after a linearly polarized beam passed through the optical spiral retarder.

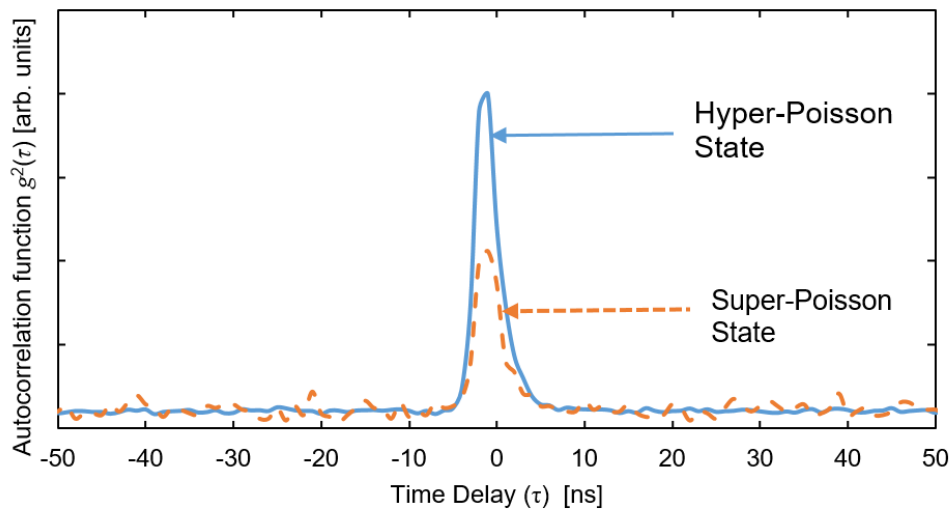

**Fig. 4.** Experimental results of $g^2(\tau)$ versus time delay ($\tau$) between signal and idler port for hyper-Poisson state and super-Poisson state.

Fig. 4 shows the experimental results of $g^2(\tau)$ as a function of time delay ($\tau$) for both cases of hyper-Poisson state and the super-Poisson states. The hyper-Poisson state exhibits a significantly higher bunching peak compared to the super-Poisson state. The sample is scanned using a motorized stage and photon counting for each point is recorded via the time correlator. By analyzing the experimental data, including the coincidence count and autocorrelation function $g^2(0)$, the final images are constructed.

## 4. Experimental results and analyses

First of all, to obtain proper coincidence windows ($\Delta t$) under the hyper-Poisson state and super-Poisson state, we measured the autocorrelation function $g^2(\tau)$ as a function of time delay ($\tau$) between channel A and B with the analyzer angles at 110° (hyper-Poisson state) and 20° (super-Poisson state). As a result, we found that a coincidence window of 15 ns is sufficient for simultaneously counting signal and idler photons both for hyper-Poisson state and super-Poisson state.

The autocorrelation function at the zero-delay, $g^2(0)$, can be determined using the following equation [24]:

$$g^2(0) = \frac{R_{AB}(0)}{R_A \cdot R_B \cdot \Delta t} \tag{1}$$

Where $R_A$, $R_B$, and $R_{AB}$ are the count rate of the beam pass with the rotating analyzer (channel A), that of the fixed analyzer (channel B), and the coincidence count rate of channels A and B, respectively. Fig. 5 shows the coincidence count and the autocorrelation function $g^2(0)$ versus the analyzer angle for different integration times, including 10, 20, 40, and 80 s. All the coincidence count rate exhibit a sinusoidal curve with respect to the analyzer angle, similar to the classical polarization measurement. However, unlike normal optical interference measurements, it should be noted that the interference-like pattern is obtained from the coincidence of photons on the two different paths, rather than the two paths of light waves interfering with each other. On the other hand, the plot of the autocorrelation function concerning the analyzer angle distinguishes between the super-Poisson state which features a sharp dip region, and the hyper-Poisson state, characterized by a flat region. From these results, it is expected that the proposed method enables us to realize an imaging technique with high robustness and high stability because it requires no real light-wave interference but only time-correlation of electrical pulse signal.

Here, we assume that the sharp dips in the super-Poisson state can be described using the Gaussian distribution function. The measurement of time-correlation experiments typically follows a Gaussian distribution due to the statistical nature of random fluctuations, arbitrariness, and timing uncertainties, which contribute to the spread of the measured experimental values. In this case, the dip is expressed with the following equation was obtained through curve-fitting analysis of the experimental data:

$$y = y_0 + A\exp\left[-4\ln2 \times \left\{\frac{(x-x_c)}{\text{FWHM}}\right\}^2\right] \qquad (2)$$

The red solid lines in $g^2(0)$ plots in Fig. 5 show the fitted curves using Eq. (2). Note that curve-fitting has been performed twice for each integration time because there are two dips in the analyzer angle range measured (0−220°). The standard deviations, which correspond to errors in measurements, obtained from the curve fits to the measured data are shown in Table 1.

As the integration time increases, the error is progressively reduced, leading to an improvement in the SNR. Meanwhile, the increase in integration time leads to an increase in total measurement time. It is suggested that the integration time of 20 s is enough to obtain moderate image contrast.

The coincidence rate exhibits a sinusoidal curve, similar to the methodology used in classical polarization microscopy. Therefore, the contrast of a coincidence rate image can be regarded as similar to a classical image. However, it is suggested that the coincidence rate image can be obtained by an extremely low power illumination. Additionally, the image contrast might be further enhanced using the autocorrelation function compared to the coincidence rate image, owing to the abrupt change in the polarization angle dependence.

Here, we defined that the maximum sensitivity and dynamic range correspond to the maximum gradient and the full width at half maximum (FWHM), respectively. For the coincidence rate, the maximum sensitivity and dynamic range are found to be $\pi/180$ [/°] and 90 [°], respectively. Maximum sensitivities and dynamic ranges obtained from curve fitting for the autocorrelation function are presented in Table 1. As shown in the table, the maximum sensitivity for autocorrelation function is several times higher than that of the coincidence rate in general, while the dynamic range is lower.

Table 1. Standard deviation, Maximum sensitivity, and dynamic range obtained from curve fits for the autocorrelation function at different integration times.

| Integration Time | Standard deviation (%) | | Maximum sensitivity [/°] | | Dynamic range [°] | |
|---|---|---|---|---|---|---|
| | 1st dip | 2nd dip | 1st dip | 2nd dip | 1st dip | 2nd dip |
| 10 s | 9.1 | 11 | 0.057 | 0.028 | 15.4 | 27 |
| 20 s | 5.4 | 7.1 | 0.037 | 0.031 | 21.1 | 26.4 |
| 40 s | 5.1 | 5.5 | 0.035 | 0.024 | 23.6 | 28.9 |
| 80 s | 5.8 | 5.2 | 0.027 | 0.028 | 25.6 | 26.5 |

As mentioned before, the depth of the dip in the autocorrelation function corresponds to the $g^2(0)$ value in the super-Poisson state. As the thermal photon generation rate increases, the $g^2(0)$ value decreases, which makes the dip more pronounced.

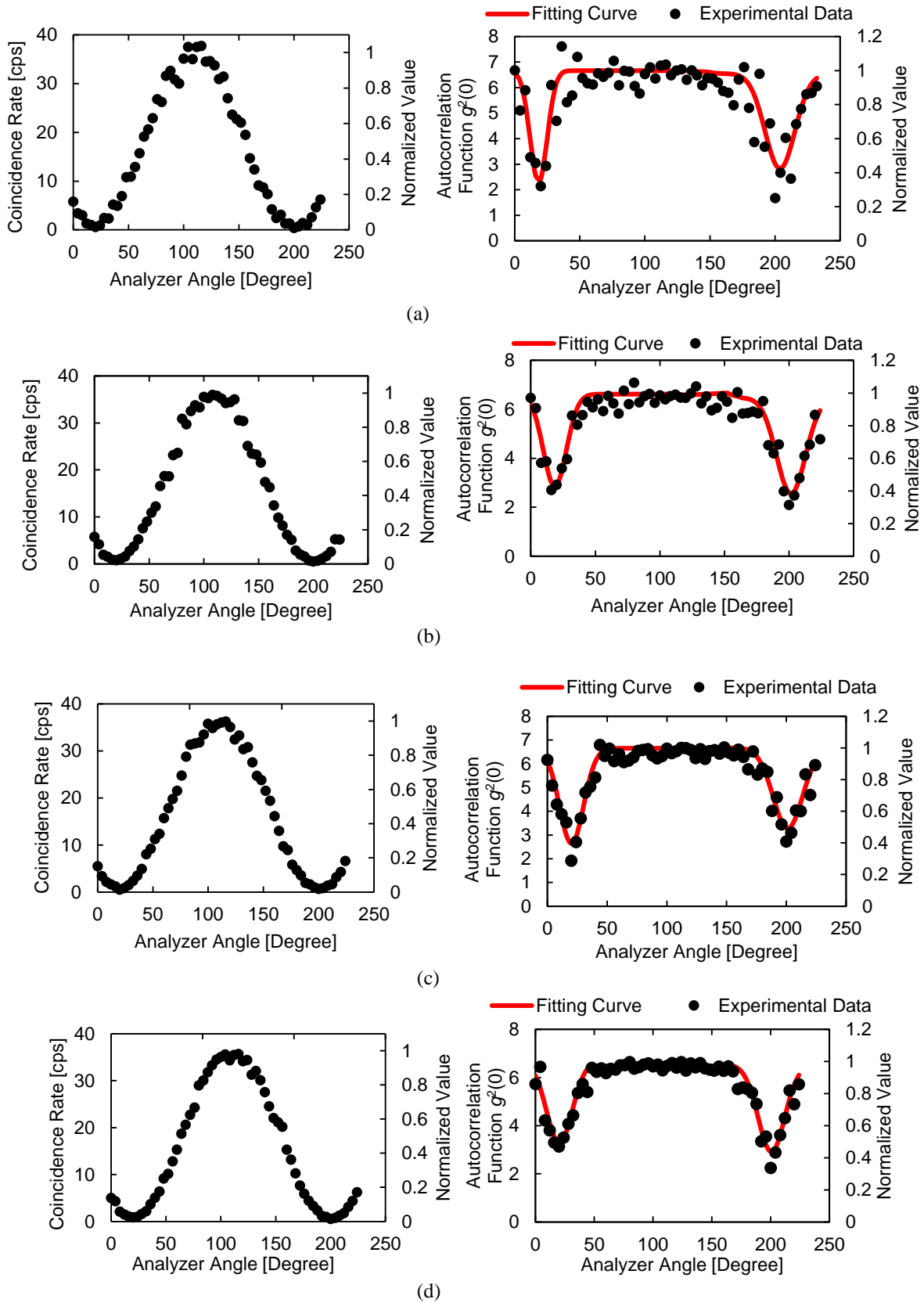

**Fig. 5.** Coincidence count and autocorrelation function at zero delay $g^2(0)$ versus analyzer angle at different integration times: (a) 10 s; (b) 20 s; (c) 40 s; (d) 80 s.

## 5. Demonstration of quantum imaging

To demonstrate our imaging method, a spiral retarder optical sample has been utilized. Fig.6 (a) presents the photo of spiral retarder optical sample with diameter of 25.4 mm, consisting of a black part and areas of free space. The dimension of image is 30×30 mm. The images are obtained using predefined experimental parameters, including integration time (20s) and coincidence window (15ns). The two-dimensional (2D) scan images of the coincidence image and the autocorrelation image are displayed in Fig. 6(b) and (c), respectively. As shown in the figures, the image derived from coincidence counts shows that image contrast is given in a sinusoidal manner according to the azimuth angle; namely, the contrast becomes stronger in areas corresponding to nodes (coincidence count at around 70 or 160° shown in Fig. 5), while it becomes weaker in areas corresponding to antinodes (coincidence count at around 25 or 115° shown in Fig. 5).

For the autocorrelation function $g^2(0)$ image, the dark area, corresponding to the dip region under the super-Poisson state reveals the strong contrast and the bright area (i.e. the flat region under the hyper-Poisson state) weak contrast. Fig. 6 (d) and (e) compare the contrast from coincidence rate with autocorrelation function $g^2(0)$ at the purple solid bars and the black solid bars, respectively, shown in Fig. 6 (b) and (c). In Fig. 6 (d), the line profile along the purple solid line exhibits greater contrast in the autocorrelation function compared to the coincidence count. Conversely, in Fig. 6(e), the line profile along the black solid line shows better contrast in the coincidence count than in the autocorrelation function. These results can be attributed to the fact that the autocorrelation function produces a sharp dip near a coincidence count of approximately zero, where the coincidence count remains relatively unchanged. In contrast, in regions where the coincidence count varies significantly, the autocorrelation function remains nearly constant. Therefore, even if the sample induces a slight polarization rotation, it is expected to obtain a high-contrast image by setting the polarization direction of the sample-transmitted light to be in the dip bottom under a super-Poisson state.

In the measurements, the optical power applied to the sample for obtaining the image was about 0.1 pJ, which is remarkably low irradiation power compared to conventional imaging techniques. Therefore, it is suggested that our proposed method can be applicable for biological systems that are easily damaged against photo-irradiation. Furthermore, a sample in a dispersive interference medium such as wet organ tissues can be clearly observed because dispersive interference can be avoided owing to the basis of quantum interference. Obviously, these advantages can pave the way for a new frontier in biomedical imaging for biological systems.

According to Ono *et al.* [17], the quantum entanglement is crucial for enhancing the signal-to-noise ratio (SNR) in phase measurements beyond the standard quantum limit (SQL). However, their approach differs significantly from our method in terms of convenience, as phase measurements typically require optical path length accuracies on the order of several tens of micrometers. In contrast, our method avoids such precise setups in favor of a more convenient and robust measurement technique. For this purpose, we employ polarization-sensing measurements rather than using phase-sensing measurements though some drawbacks such as accuracy might come.

To compare our proposed method with existing work, the systematic analysis of signal-to-noise ratio in bipartite is defined as the ratio of the "mean contrast" to its standard deviation (mean fluctuation) [25]. The SNR value for the proposed method can be derived using the following equation [3].

$$\text{SNR}_s = \frac{|\langle s_{\text{in}} - s_{\text{out}} \rangle|}{\sqrt{\langle \delta^2(s_{\text{in}} - s_{\text{out}}) \rangle}} \qquad (3)$$

Where $S_{\text{in}}$ represents the average photon count rate at the signal port section before inserting the sample, $S_{\text{out}}$ is the average photon count rate at the signal port section after inserting the sample, and $\delta^2(S_{\text{in}} - S_{\text{out}})$ represents the statistical fluctuation or uncertainty in the difference between $S_{\text{in}}$ and $S_{\text{out}}$. Consequently, the value of SNR for the present experiments was found to be 19.7, which is comparable to the SNR value reported in the Ref. [5].

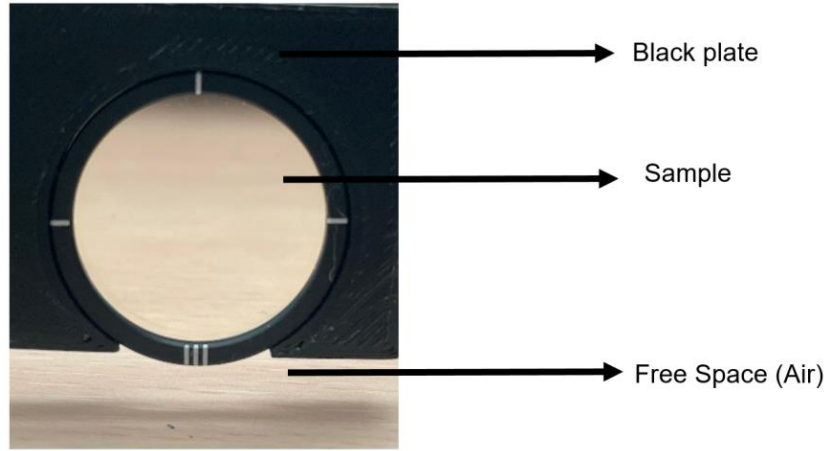

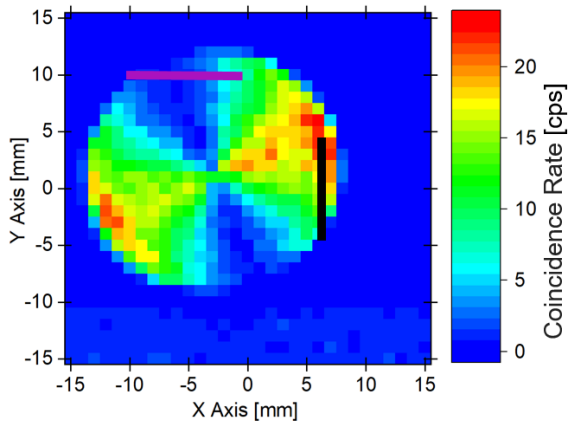
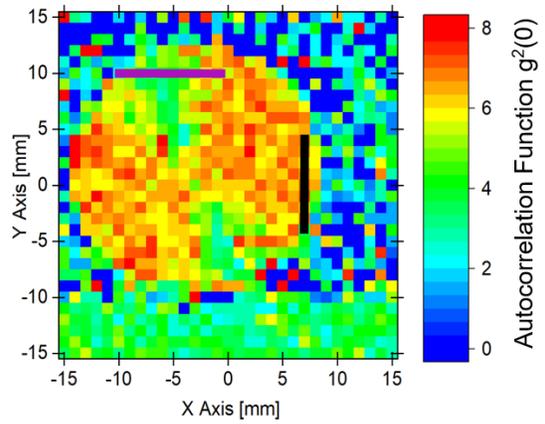

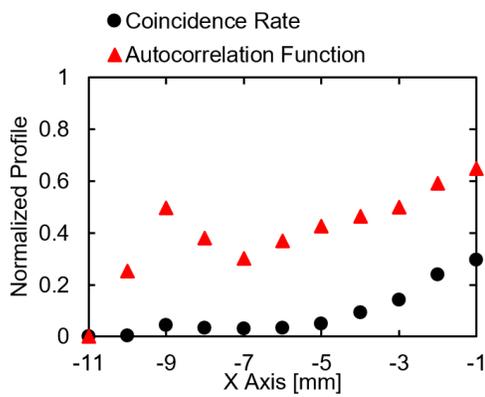
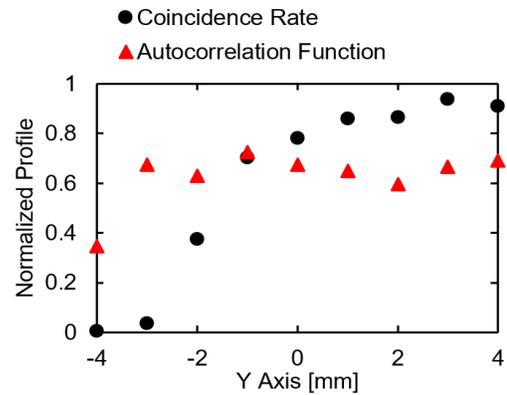

**Fig. 6.** (a) Photograph of the optical spiral retarder sample (Diameter of sample: 25.4 mm) (b) Coincidence image of spiral retarder; (c) Autocorrelation Image of spiral retarder; (d) One-dimensional line-scanned profile highlighted in purple solid line from the coincidence rate and autocorrelation function image; (e) One-dimensional line-scanned profile highlighted in black solid line from the coincidence rate and autocorrelation function image.

## 6. Conclusion

In summary, we proposed a new polarization microscopic method, in which the switching super-Poisson state and hyper-Poisson state of entangled-photons by polarization angle variation can significantly enhance the image contrast in microscopy. Our method was demonstrated through the observation of an optical spiral retarder sample. We revealed that this imaging technique is enabled by a quantum-source, which does not rely on classical optical coherence, offering substantial improvements over classical polarization imaging techniques in terms of robustness and stability.

Furthermore, this approach mitigates the susceptibility to photo- damage and photo-bleaching often associated with high-intensity laser beams. This technological breakthrough promises to expand new horizons in exploring the mechanical properties of live biological systems. However, the fluctuation noise is suggested as the current issue of our imaging technique, resulting in the increase of measurement time. In addition, although the entangled-photon beam was not focused at present, it should be focused with the lens to improve the spatial resolution, S/N ratio, and so on, but the throughput of photon might be reduced. In conclusion, we anticipate that quantum polarization microscopy, utilizing a quantum-entangled photon source, will soon advance toward practical applications in biological imaging and sensing beyond the classical limit, if further improvements such as reduction of noise and measurement time would be achieved.

**Acknowledgements**

Part of this study was supported by the JSPS KAKENHI (Grant No. 22K18292, 23K22787) and JST A-STEP (Grant No. JPMJTR22RD).

**Corresponding Author:** * samad.m.128@ms.saitama-u.ac.jp; yasuto@mail.saitama-u.ac.jp

**Data availability:** The data that support the finding of this study are available from the corresponding author upon reasonable request.

# References

1. B.M. Gilaberte, F. Setzpfandt, F. Steinlechner, E. Beckert, T. Pertsch, M. Gräfe, Perspectives for applications of quantum imaging. Laser & Photonics Reviews 13 (2019) 1900097:1-24.
2. P.A. Morris, R.S. Aspden, J.E. Bell, R.W. Boyd, M.J. Padgett, Imaging with a small number of photons. Nature Communications 6 (2015) 5913:1-6.
3. R. Tenne, U. Rossman, B. Rephael, Y. Israel, A.K. Ptaszek, R. Lapkiewicz, D. Oron, Super-resolution enhancement by quantum image scanning microscopy. Nature Photonics 13 (2019) 116-122.
4. 4. I. Kviatkovsky, H.M. Chrzanowski, E.G. Avery, H. Bartolomaeus, S. Ramelow, Microscopy with undetected photons in the mid-infrared. Science Advances 6:42 (2020) 1-6.
5. H. Defienne, B. Ndagano, A. Lyons, D. Faccio, Polarization entanglement-enabled quantum holography. Nature Physics 17 (2021) 591-597.
6. C.A. Casacio, L.S. Madsen, A. Terrasson, M. Waleed, K. Barnscheidt, B. Hage, W.P. Bowen, Quantum-enhanced nonlinear microscopy. Nature 594 (2021) 201-206.
7. P.A. Moreau, E. Toninelli, T. Gregory, M.J. Padgett, Imaging with quantum states of light. Nature Reviews Physics 1 (2019) 367-380.
8. Y. Shih. Entangled biphoton source-property and preparation. Rep. Prog. Physics 66 (2003) 1009–1044.
9. R. Horodecki, P. Horodecki, M. Horodecki, K. Horodecki. Quantum entanglement. Rev. Modern Physics 81 (2009) 865-942.
10. R.D. Boyd, Nonlinear Optics, 3rd Edition, Elsevier Science: Oxford, UK, (2008) pp. 1–59.
11. C. Couteau, Spontaneous parametric down-conversion. Contemporary Physics 59 (2018) 291-304.
12. M.H. Rubin, Y. Shih, Resolution of ghost imaging for no degenerate spontaneous parametric down-conversion. PHYSICAL REVIEW A 78 (2008) 033836:1-7.
13. S. Karmakar, Y. Shih, Two-color ghost imaging with enhanced angular resolving power. PHYSICAL REVIEW A 81 (2010) 033845:1-6.
14. G.B. Lemos, V. Borish, G.D. Cole, S, Ramelow, R. Lapkiewicz, A. Zeilinger, Quantum imaging with undetected photons. Nature 512 (2014) 409-412.
15. D.A. Kalashnikov, A.V. Paterova, S.P. Kulik, L.A. Krivitsky, Infrared spectroscopy with visible light. Nature Photonics 10 (2016) 98-101.
16. P.G. Kwiat, K. Mattle, H. Weinfurter, A. Zeilinger, A.V. Sergienko, Y. Shih, New high-intensity source of polarization-entangled photon pairs. Physical Review Letter 75 (1995) 4337- 4341.
17. T. Ono, R. Okamoto, S. Takeuchi, An entanglement-enhanced microscope. Nature Communications 4:2426 (2013) 1-7.
18. T. Nagata, R. Okamoto, J.L. O'brien, K. Sasaki, S. Takeuchi, Beating the standard quantum limit with four-entangled photons. Science 316 (2007) 726-729.
19. V. Giovannetti, S. Lloyd, L. Maccone, Quantum metrology. Physical Review Letter 96 (2006) 010401:1-4.
20. Y. Zhang, Z. He, X. Tong, D.C. Garrett, R. Cao, L.V. Wang, Quantum imaging of biological organisms through spatial and polarization entanglement. Science Advances 10:10 (2024) 1-11.
21. Z. He, Y. Zhang, X. Tong, L. Li, L.V. Wang, Quantum microscopy of cells at the Heisenberg limit. Nature Communications 14 (2023) 2441:1-8.
22. Correlated photon-pair source. Available online https://www.thorlabs.com (29th July, 2024).
23. S. David, J. Gregg, Quantum Metrology, Imaging, and Communication. Springer: 2018. [https://doi.org/10.1007/978-3-319-46551-7].
24. EDU-QOP1(/M) Quantum Optics Kit. Available online: https://www.thorlabs.com (9th April, 2024).
25. G. Brida, M.V. Chekhova, G.A. Fornaro, M. Genovese, E.D. Lopaeva, I.R. Berchera, Systematic analysis of signal-to-noise ratio in bipartite ghost imaging with classical and quantum light. Physical Review A—Atomic, Molecular, and Optical Physics 83 (2011) 063807: 1-10.